\documentclass[letterpaper]{JHEP}
\usepackage[dvips]{epsfig}
\usepackage{epsfig}
\usepackage{graphicx}

\title{Inflating $p$-branes}
\author{Ruth Gregory\\ Centre for Particle Theory, 
University of Durham,\\ South Road, Durham, DH1 3LE, U.K.}

\abstract{
We look for solutions in Einstein gravity corresponding to inflating
braneworlds of arbitrary dimension and co-dimension. These solutions
correspond to isolated sources (no long range fields). Using dynamical
systems techniques, we show that there exists a unique solution corresponding
to a black $p$-brane with a regular horizon at the location of the brane.
The solution is {\it not} however asymptotically flat, but has global
deficit angles.
}

\keywords{supergravity solutions, black holes, $p$-branes}
\preprint{DCPT-03/23, hep-th/0304262}

\def\ie{{\it i.e.,}}

\newcommand{\be}{\begin{equation}}
\newcommand{\ee}{\end{equation}}
\newcommand{\bea}{\begin{eqnarray}}
\newcommand{\eea}{\end{eqnarray}}
\newcommand{\bml}{\begin{mathletters}}
\newcommand{\eml}{\end{mathletters}}

\begin{document}

\section{Introduction}

Inflation is one of the key tenets of the modern standard cosmological
model. It allegedly solves many of the problems of naturalness in the
old hot big bang model, and perhaps it greatest allure is in `predicting'
a scale invariant perturbation spectrum -- in ever increasing agreement
with the observations of inhomogeneities in the microwave background
\cite{MAP}.
It is however telling that no definitive inflationary model exists,
indeed, no definitive observation of even a single scalar particle has
yet been made, let alone a multitude, as many inflationary models require.

Although the issue of exponential expansion in the early 
universe is nominally an open question, current observations 
do seem to indicate that the universe is once more in a stage 
of gentle accelerated expansion \cite{Aexp}. 
Assuming that gravity is approximately four-dimensional and 
Newtonian/Einsteinian at large scales indicates that our universe
has a negative equation of state at the current time \cite{negw}.
This inevitability of an accelerating universe places
strong demands on any underlying theory of fundamental
physics -- whether it be simply the challenge to produce
an appropriate equation of state \cite{Quint}, or, more subtle questions
about the consistency of a de-Sitter asymptotic state for
the universe \cite{dSprob}.

Braneworlds \cite{BW} are an interesting orthogonal development
in the attempt to describe our universe within the context
of a fundamental higher energy theory. The braneworld scenario
imagines us as being confined to a four-dimensional
hyperplane in a higher dimensional spacetime.
Standard Model interactions and most usual phsyics
is confined to the brane universe, with
only gravity (or some small number of zero-mode fields) 
propagating in the bulk.
Such a set-up can provide an interesting alternate solution
to the hierarchy problem \cite{ADD,RS1}, but it is crucial
that such a framework is gravitationally and cosmologically consistent.
Of course, on small scales we can allow KK graviton modes to be present, 
but on larger scales gravity must be Einsteinian to be consistent with
observation. Of course, on the extremely large scale, it is possible
that gravity is not in fact Newtonian (a possibility explored
by Milgrom \cite{MILG}), and in fact an 
unexpected bonus of some braneworld models is that gravity can be 
modified on large scales \cite{MODG} in a generally covariant fashion.

Within the context of Randall-Sundrum (RS) braneworlds \cite{RS1,RS2}, a 
phenomenologically motivated model rather similar to
heterotic M-theory compactifications \cite{LOSW}, in which there is
only one extra dimension, the description of our universe
is particularly simple: Our universe is simply a hyperplane
living in five negatively curved dimensions. Perturbative gravity
is easily shown to be identical to perturbative four dimensional
Einstein gravity \cite{RS2,GT}, and in the pure RS
set-up where only gravity propagates in the bulk, the
full set of solutions for our cosmological braneworld
are known and easily found as moving branes in a Schwarzschild
anti-de Sitter (adS) bulk \cite{BCG}. In particular, the standard
inflationary universe is simply a uniformly moving brane in
pure adS spacetime.

In string theory however, we live in more than five dimensions, and
even in the heterotic M-theory compactification \cite{HetM}, the
universe is only effectively five-dimensional in a small range of
high energies. The more string-motivated Arkani-Hamed et.\ al.\
compactifications \cite{ADD} have many extra dimensions.
However, many of these models do not include the gravitational
effect of the energy-momentum of the brane itself.
One of the reasons that the RS models are so easy to deal with
is the fact that gravity in one spatial dimension is trivial
(recall that the universe is homogeneous and isotropic to leading
order), and this
renders the problem of finding cosmological solutions straightforward.
With more extra dimensions however, the effect of consistently
including the stress-energy of the brane becomes nontrivial.
If we have three or more extra spatial dimensions, local warping of those
extra dimensions induced by the stress-energy of the braneworld
creates a naked null singularity at the putative brane \cite{CPB}, which
although it does not stop physics in the bulk being well-defined,
does mean that any physical description of the brane is
highly dependent on the way the brane is modelled \cite{CRR}.
(A particularly interesting variant being the nonsingular ``blown-up''
$p$-brane \cite{ELL}.)

Given the natural interest in inflationary solutions for our
universe, and the interest in finding braneworld resolutions
to various cosmological problems, it is an obvious question to
try to find general inflating braneworld solutions. Of course,
these are well known for the RS case of codimension one. However,
inflating braneworlds for higher codimension are not explicitly
known. (Although a recent paper by Olasagasti and Tamvakis \cite{OT}
looks for inflating solutions exterior to a global defect, 
extended to include the core by Cho and Vilenkin \cite{CV}.) 
The existence of asymptotically flat solutions has been
assumed in work attempting to incorporate stringy inflation via
brane motion on the compactification manifold \cite{bbar}, and
the existence of regular solutions assumed in work
seeking a self-tuning mechanism for the small value of the
cosmological constant \cite{Dlam}. The fact remains however, that
there are no isolated inflating braneworld-type solutions known for 
codimension three or higher.

To understand why this might be a nontrivial question, consider
the marginal case of codimension two. In many six-dimensional
braneworld models, our brane appears as a conical deficit
with an induced Minkowski flat metric.
This is easily modelled within field theory as a `cosmic string'
defect, just as the RS model is a domain wall. However, there
are two sorts of field theory vortex -- the local vortex, which
has a conical deficit as above, or the global, which has a long-range
Goldstone boson field, and is not simply an isolated conical deficit. 
The gravitational effect of this long-range field is to cause
a self-compactification of the spacetime \cite{GS} in a manner similar
to that of the vacuum domain wall \cite{GWG}, and the induced metric of
our braneworld is in fact a de Sitter universe \cite{G1} again like
that of the domain wall \cite{VK}. If a Minkowski braneworld
metric is desired, then it is necessary to introduce a negative
cosmological constant \cite{G2}, and one recovers a hierarchy
resolving RS-style model \cite{G2,CK}. The interpretation
of  this gravitational interaction is that if one forces a braneworld
to have a particular induced metric, say Minkowski, then the
solution of the gravitational equations will generically be
singular at a finite distance from the braneworld \cite{CK,CKS}. 
However, if one tunes a bulk cosmological constant against a 
braneworld Hubble expansion, then there is a one parameter 
family $H(\Lambda)$ for which a nonsingular solution exists \cite{GS}. 
This would appear to be a general result, not just confined to the 
global vortex in Einstein gravity, as Berglund et.\ al.\ \cite{DSS} 
found a similar 
behaviour within low energy string gravity while looking for 
inflating codimension two solutions in an attempt to incorporate 
de Sitter space into string theory in a natural way. 
It is tempting to conjecture that the singularity of the 
self-gravitating cosmic $p$-brane \cite{CPB} could also be
resolved by a similar process, however, there are two key
differences with the global vortex: The $p$-brane singularity
lies at its core, rather than at finite distance, and
is asymptotically flat -- \ie\ not compact. There is also
the lack of energy-momentum in the bulk, since there is no long
range Goldstone field. It is therefore not at all clear that 
allowing the brane to inflate will solve the problem of the 
singularity at the core -- or that if it does, another singularity
might not appear at finite distance.

Returning to the isolated codimension two inflating brane, note that
this will be a solution of the Einstein equations where the metric 
can be chosen to take the form
\be
ds^2 = A^2(r) \left [ dt^2 - \cosh^2 t \ d\Omega^2_{D-3} \right ]
- B^{-2}(r) dr^2 - B^2(r) d\theta^2
\ee
Fortunately, we do not need to actually write down the Einstein
equations and solve them, since if we double analytically continue
this metric by making $\theta$ a time coordinate, and $t\to i\chi + \pi/2$
a spacelike coordinate, this is readily seen to be a spherically 
symmetric static solution in $D$ dimensions, and is hence the
Schwarzschild solution. Thus
\be
\label{kkinsmet}
ds^2 = r^2 \left [ dt^2 - \cosh^2 t \ d\Omega^2_{D-3} \right ]
- \left ( 1 - \left ( {r_+\over r} \right)^{D-3} \right )^{-1} dr^2
- \left ( 1 - \left ( {r_+\over r} \right)^{D-3} \right ) d\theta^2
\ee
Although the transverse (braneworld) dimensions are those of
a lorentzian inflating universe, this is easily seen to have
the $(r,\theta)$ geometry of the euclidean black hole `cigar'.
As such, $r=r_+$ can be thought of as the location of the brane.
In fact, the metric (\ref{kkinsmet}) with $D=5$ was used 
by Witten \cite{WIT} to demonstrate an instability of the KK vacuum.

Conventionally, in black hole thermodynamics, the periodicity of
Euclidean time (here the $\theta$-angle) is fixed by requiring 
regularity at $r=r_+$, however, if we are looking for a solution
corresponding to a codimension two brane, we do not want
regularity, rather, it is precisely the conical deficit at
$r=r_+$ that will indicate the presence of the brane. Just
as with the standard cosmic string, there should be a deficit
of $8\pi\mu$ where $\mu$ is the energy per unit $p$-area of
the brane in Planck units. Computing the metric near $r=r_+$
gives the relation between
the periodicity of the $\theta$-angle, the energy of
the brane, and $r_+$  as:
\be
\delta\theta = {4\pi r_+\over D-3} \left ( 1 - 4G\mu \right)
\ee
As $r\to\infty$, the metric is asymptotically the 
KK vacuum ${\bf R}_{p+2}\times S^1$, written in Rindler
coordinates (${\bf X}_{p+1} = r\cosh t \ {\bf n}_{p+1}$, 
$T = r\sinh t$), with
the internal circle having dimension $(D-3)/[2r_+(1-4G\mu)]$.

Although this solution is regular, it is different in 
character from the inflating wall solution, which is simply
a moving brane in some five-dimensional background bulk.
Here, the bulk is necessarily the KK vacuum, rather than
being noncompact, and a {\it `bubble of nothing'} is present in
the spacetime. If already the spacetime of an inflating codimension
two brane is so phenomenologically different from codimension one,
we cannot expect to use intuition to deduce what higher
codimension inflating branes will look like. We must therefore
actually search for solutions, which is what we will now do. 
We first derive the Einstein equations, and revisit the Poincar\'e
invariant $p$-branes of reference \cite{CPB} in the context of a
dynamical system. Then we analyse the inflating brane solutions.
We comment on branes with anti-de Sitter geometries before concluding.

\section{Gravitational Equations}

In general, a $p$-brane produced by some localized source
with energy and tension of the same magnitude need not be
Poincar\'e invariant, but can in fact have an induced metric
which is constant curvature. For an inflating brane, this
will be constant positive curvature. We therefore look for
a solution which is a warped product of this constant curvature
worldbrane metric, and dependent on some orthogonal coordinate.
The metric can be written as:
\be
ds^2 = A^2(r) \left [ dt^2 - \cosh^2 (\sqrt{\kappa}t) \, 
d\Omega_p^{\prime2} \right ] - B^2(r) dr^2 - C^2(r) d\Omega_n^2
\label{pbrane}
\ee
where $\kappa$ has been added explicitly for comparison with the
standard Poincar\'e invariant cosmic $p$-branes. Note this metric
has brane dimension $p+1$ and codimension $n+1$.

The Einstein equations for this metric are:
\bea
R^t_t &=& {1\over B^2} \left [ {A''\over A} - {A'B'\over AB}
+ p {A^{\prime2} \over A^2} + n {A'C'\over AC} \right ] - {p\kappa\over A^2}
\label{rtt}\\
R^r_r &=& {(p+1)\over B^2} \left ( {A''\over A} - {A'B'\over AB} \right )
+ {n\over B^2} \left ( {C''\over C} - {C'B'\over CB} \right ) 
\label{rrr}\\
R^\theta_\theta &=& {1\over B^2} \left [ {C''\over C} - {C'B'\over CB}
+ (p+1) {A'C'\over AC} + (n-1) {C^{\prime2}\over C^2} \right ]
-{(n-1)\over C^2} \label{rtheta}
\eea
At this point, it is immediate that there is no possibility of
a {\it `bubble of nothing'} type of solution to the higher codimension
brane with $A\sim r$, and $B,C$ roughly constant, since (\ref{rtheta})
cannot satisfy the Einstein equation $R^\theta_\theta=0$. 

In the case of the Poincar\'e invariant brane, where there is no
$\kappa$ term in (\ref{rtt}), the Einstein equations can be directly 
integrated for a suitable choice of the function $B$, giving the 
cosmic $p$-branes \cite{CPB}
\be
\label{cpbmet}
ds^2 = \left ( 1 - ({r_+\over r})^{n-1} \right ) ^a \left[ dt^2 - dy_p^2\right]
- \left ( 1 - ({r_+\over r})^{n-1} \right ) ^b dr^2
- r^2 \left ( 1 - ({r_+\over r})^{n-1} \right ) ^c d\Omega_n^2
\ee
where
\be
a = {\sqrt{n}\over \sqrt{(n+p)(p+1)}} \;, \qquad 
b = -{[(n-2)+a(p+1)]\over (n-1)} \;,\qquad 
c = 1+b
\ee

For the inflating brane we have not been able to find an exact
analytic solution, however, by re-expressing the Einstein
equations as a two-dimensional dynamical system, it is possible
to demonstrate the existence of a solution, and to derive its
general form.

To do this, let $B\equiv A$ and define
\bea
X &=& {A'\over A} +{n\over p} {C'\over C} \label{xdef}\\
Y &=& {C'\over C} \label{ydef}
\eea
in which case (\ref{rtt} - \ref{rtheta}) can be rewritten as
\bea
X' &=& X^2 - \kappa - {n(n+p)\over p^2} Y^2 \label{xprime} \\
Y' &=& {p(p+1)\over n} (X^2-\kappa) - pXY - {(n+p)\over p} Y^2 \label{yprime}
\eea
together with the constraint
\be
\label{constr}
{\cal C} (X,Y) = p(p+1) (X^2-\kappa) - {n(n+p) \over p} Y^2 
= n(n-1) {A^2\over C^2} \;.
\ee
Note that 
\be
{d{\cal C}\over dr} = 2{\cal C} \left [ X - {(n+p)\over p} Y \right ]
\ee
hence ${\cal C}=0$ represents an invariant hyperboloid in the phase plane.

In addition, there are two pairs of critical points
\bea
P_{\pm}&=&\pm\sqrt{\kappa} \left(1 , 0\right)\\
Q_{\pm}&=&\pm\sqrt{\kappa} \left(\sqrt{n+p\over p}, \sqrt{p\over n+p}\right)
\eea
for $\kappa=1$, which merge to a single critical point at the origin
for the Poincar\'e brane, $\kappa=0$.

Although we will give a more detailed analysis of the inflating brane
phase plane in the next section, the main features to note at this
stage are that $P_\pm$ lie on the invariant hyperboloid ${\cal C}=0$, 
whereas the $Q_\pm$ have ${\cal C}=np\kappa>0$; both are therefore in
the physically allowed region of the phase plane ${\cal C}\geq0$.
The $P_\pm$ are saddle points, but the precise nature of the $Q_\pm$
critical points depends on the overall dimensionality of
spacetime. In general $Q_+$ is an attractor, and $Q_-$ is
a repeller, however, if $D<10$, these critical points are foci,
hence trajectories approach or repel in a vortical fashion.

Before turning to the inflating brane solutions, it is actually useful
to first analyse the Poincar\'e phase plane, since in this case
we actually have the metric explicitly, and can therefore directly calculate
$X$ and $Y$. The phase plane is given by solving (\ref{xprime},
\ref{yprime}) with $\kappa=0$. As already mentioned, the critical
points merge into a single degenerate critical point at the origin:
${\bf P} = (0,0)$. Clearly the invariant curve $\cal C$
is now a pair of straight lines $Y = \pm \gamma^{-1} X$, where
we have defined 
\be
\label{gammadef}
\gamma = {\sqrt{n(n+p)/(p+1)}\over p}
\ee
for later use.
The phase plane is shown in figure \ref{fig:poin32} for the
same values $n=2$ and $p=3$ for comparison with figure \ref{fig:ds32}.
\FIGURE{
\includegraphics[height=15cm]{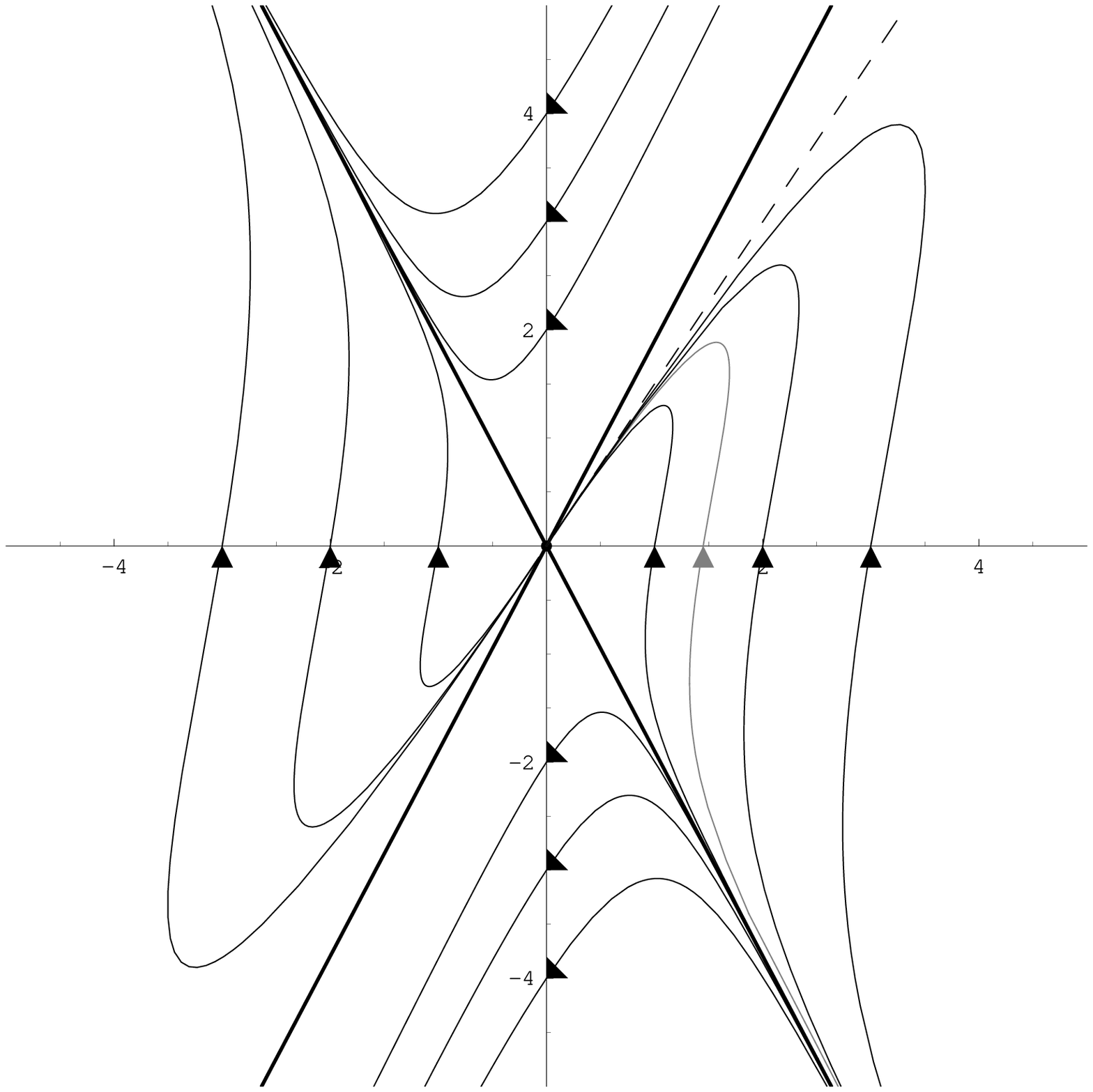}
\caption{The phase plane of the Poincar\'e $p$-brane. The invariant
curve ${\cal C}=0$ is the pair of thick stright lines, and the Minkowski
spacetime solution is the dotted line. A generic $p$-brane solution is shown
in grey. }
\label{fig:poin32}
}

Notice that a typical trajectory in the right hand quadrant starts off
at large $(X,-Y)$, asymptoting ${\cal C}=0$, curves around, and approaches
{\bf P} tangent to the dotted line $Y=pX/n$ in general. Solving for the
asymptotic small $r$ region gives
\be
X = {1\over pr} \qquad,\qquad Y = -{1\over p\gamma r}
\ee
which in turn gives the solution
\be
A \propto r^{{1\over p}+{n\over p^2\gamma}} \qquad,\qquad
C \propto r^{-1/p\gamma}
\ee
recalling that $B=A$ in these coordinates, and transforming to the
radial coordinate in which the $p$-brane metric (\ref{cpbmet}) takes
its canonical form ($A^{-2} dr_p = Adr$) indeed shows that this is
the ``near-core'' r\'egime $r\to r_+$ of the $p$-brane (which of
course is a null singularity). For 
$\{X,Y\} \to {\bf P}$ on the other hand, we have
\be
A \sim 1 - {\alpha_0\over r^{n-1}} \qquad,\qquad
C = r \left ( 1 + {(p+2)\alpha_0\over (n-2)r^{n-1}} \right)
\ee
the asymptotic far-field r\'egime of the $p$-brane solution.

Specifically, the exact form of the Poincar\'e $p$-brane solution 
(\ref{cpbmet}) gives
\bea
X &=& {2u\over3r_+} (1-u)^{\sqrt{5\over8}-1}\left [ 
1 - {(\sqrt{5}+2\sqrt{2})\over4\sqrt{2}}u \right ] \\
Y &=& {u\over r_+} (1-u)^{\sqrt{5\over8}-1} \left [ 
1 - {(\sqrt{5}+2\sqrt{2})\over2\sqrt{5}}u \right ]
\eea
We see therefore that altering the mass of the solution simply scales
the plot in the phase plane. A representative trajectory with $r_+=4/30$ 
is shown in grey in figure \ref{fig:poin32}. Note that for $r_+=0$, \ie\ flat
space, we have $Y=pX/n$ (shown as a dotted line), which is a separatrix
in the Poincar\'e phase plane. This also shows manifestly that the
solutions are asymptotically flat.

\section{Inflating branes and their global structure}

Now turn to the inflating brane phase plane. 
A phase plot of this system for the values $p=3$, $n=2$ is shown in
figure \ref{fig:ds32}. For large $(X,Y)$, the system asymptotes the
Poincar\'e plane, therefore we expect the physical solutions to
correspond to trajectories in the right hand exterior of the
invariant hyperboloid. Indeed, comparing figure \ref{fig:ds32} to
figure \ref{fig:poin32},
we spot that there are a similar family of trajectories
asymptoting the invariant hyperboloid for large $X\propto-Y$
which now terminate on $Q_+$. However, there are now some
additional interesting solutions. The splitting of ${\bf P}$ 
into the two pairs of critical points allows a single trajectory
from $P_+$ to $Q_+$. Also of later use is the existence of the
other stable manifold trajectory connecting large negative $X\propto Y$
to $P_+$.
\FIGURE{
\includegraphics[height=15cm]{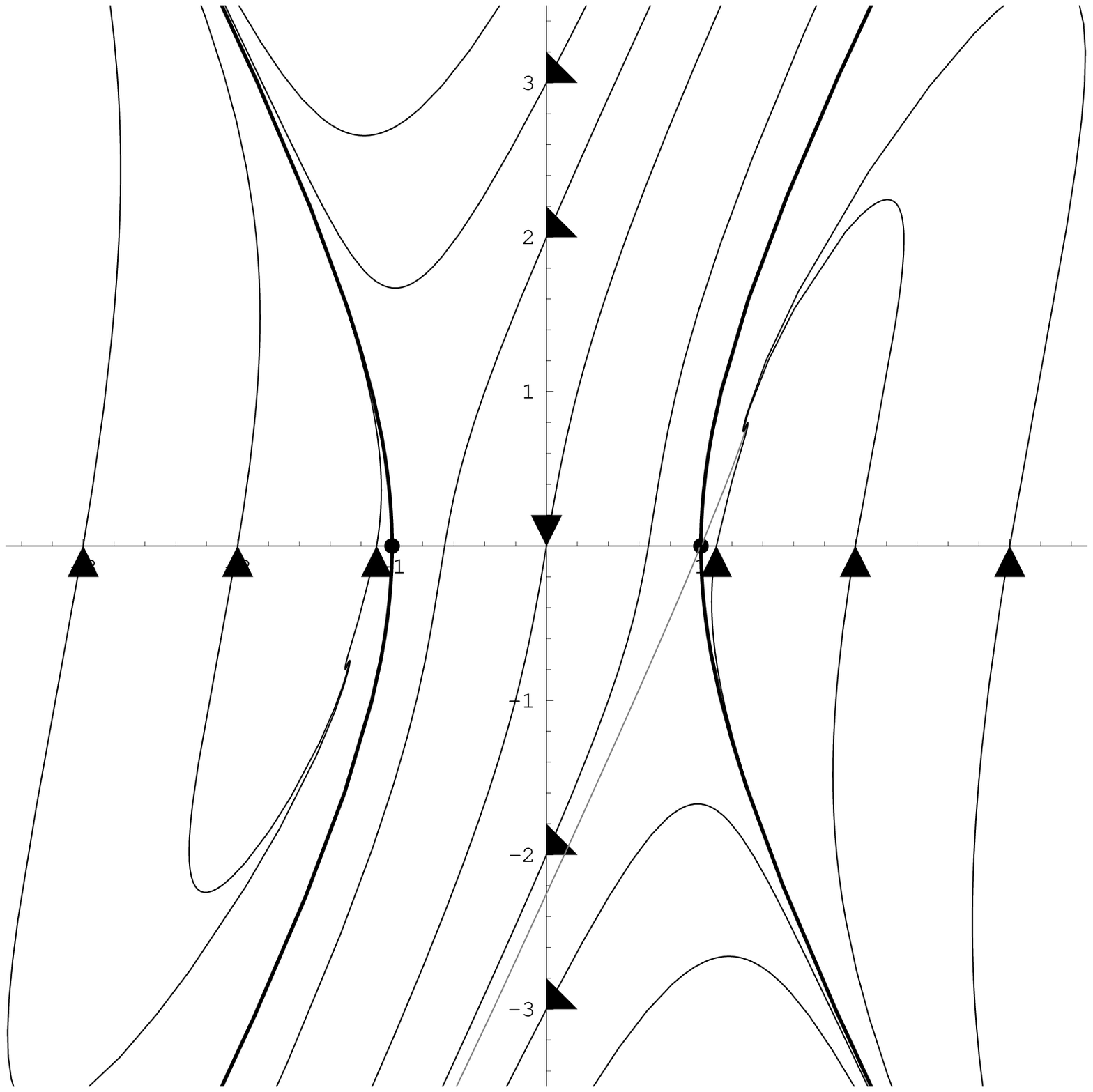}
\caption{The inflating brane phase plane. The invariant hyperboloid
${\cal C}=0$ is shown in bold again, the critical points by a dot,
and the candidate  trajectory from $P_+$ to $Q_+$ in grey (as well
as its continuation into the interior-horizon region).}
\label{fig:ds32}
}

Dealing with the typical trajectory first, we note that these are
analogous to the Poincar\'e curves - they asymptote the near
singularity r\'egime of (\ref{cpbmet}). However, unlike the Poincar\'e 
solutions, these cannot have an asymptotically flat solution. Apart
from ${\cal C}=0$ itself, all of these trajectories terminate on $Q_+$.
An analysis of $Q_+$ shows that the metric in this asymptotic region is
\be
ds^2 \simeq {p\over(p+n)} \rho^2 \left [ dt^2 - \cosh^2 t \ d\Omega_p^2 \right]
-d\rho^2 - {(n-1)\over(n+p)} \rho^2 d\Omega_n^2 \label{qplussol}
\ee
as $\rho \to \infty$. This latter metric, while asymptotically locally
flat, is not asymptotically flat, as it has global deficit angles in the
spatial $S^n$ part of the metric, as well as in the inflating braneworld
part.

The net result is that apart from this asymptotic global deficit angle,
the metric of these solutions is somewhat similar to the Poincar\'e 
$p$-brane, in that it has a null naked singularity as a near-field
limit, which has an infinite area. This singularity integrates out to
an ALF spacetime with a global deficit angle. Presumably, as with the
standard $p$-brane \cite{CRR}, propagators are well defined on this spacetime,
although we have not explored this issue.

One might therefore think that the inflating brane spacetime is similar
to that of the Poincar\'e brane, however,
there is one other possibility not present in the Poincar\'e phase plane, 
illustrated as the grey trajectory in figure \ref{fig:ds32}, 
and that is the trajectory from $P_+$ to $Q_+$. Analysing the spacetime
near $P_\pm$ shows that these critical points correspond to horizons;
in a suitable coordinate system, the metric for the solution
near $P_+$ is 
\be
\label{nrpmet}
ds^2 \simeq \rho^2 \left ( 1 - {n(n-1)\over 3(p+1)(p+2)C_0^2} \rho^2 \right )
dx^2_{p+1} - d\rho^2 - 
C_0^2 \left ( 1 + {(n-1)\rho^2\over (p+2)C_0^2} \right ) d\Omega_n^2
\ee
as $\rho\to0$. 

To see this is a simple horizon, and also to obtain the maximal analytic
extension of the spacetime, return to the phase plane coordinate $r=-\ln\rho$,
and let
\be
\label{krusk}
U = e^{t-r} \qquad,\qquad V = - e^{-t-r}
\ee
in which the near $P_+$ metric (\ref{nrpmet}) is
\be
ds^2 \sim dUdV - {(U-V)^2\over 4} d\Omega_p^2 - C_0^2 d\Omega_n^2
\ee
as with conventional Kruskal coordinates (see figure \ref{fig:krs})
\FIGURE{
\includegraphics[height=15cm]{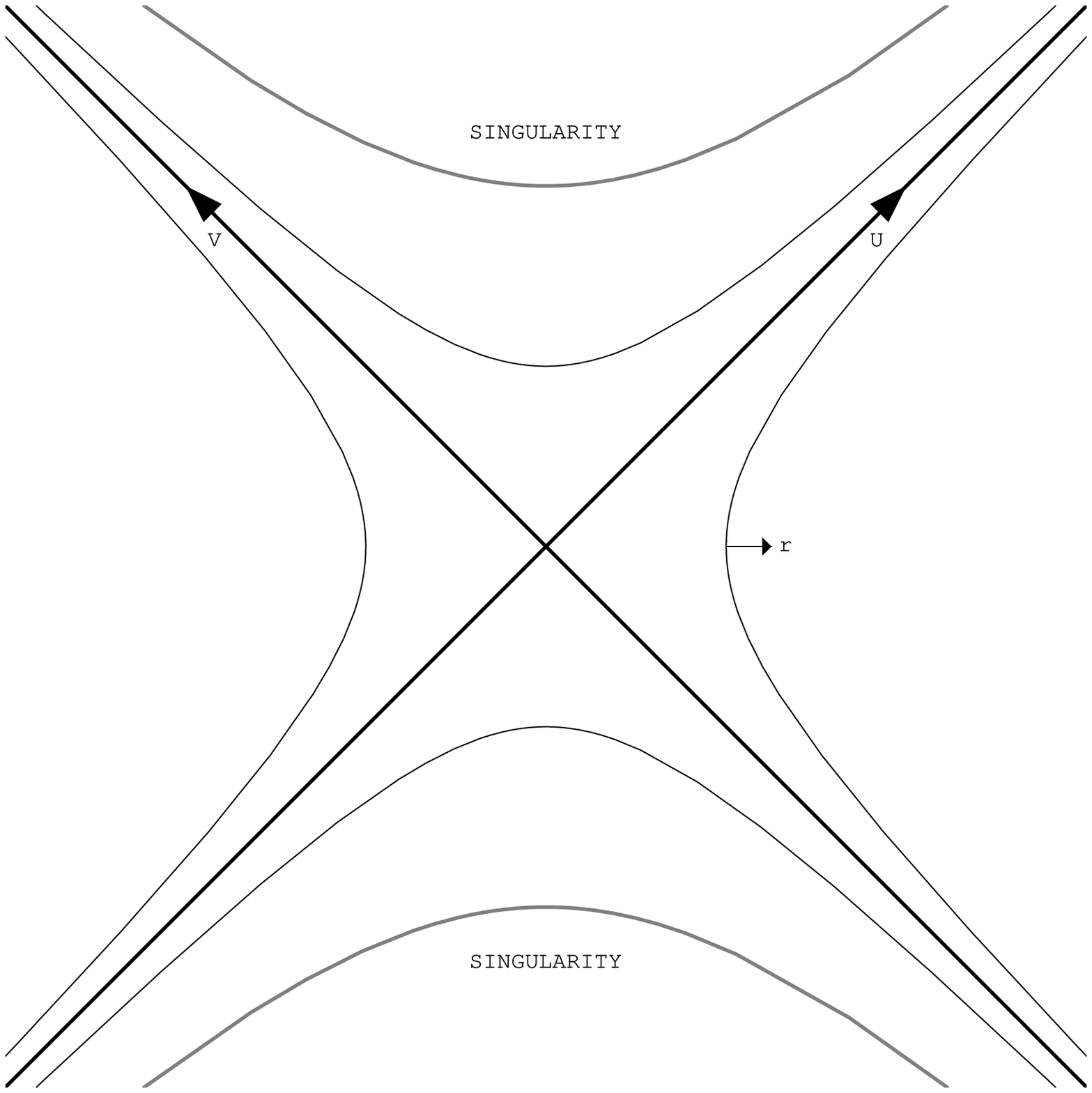}
\caption{The Kruskal diagram of the inflating brane. The horizon (shown
as the thick lines) is represented by the critical point $P_+$. The
$r$-coordinate is shown in the exterior region, and the singularity
as a grey line.}
\label{fig:krs}
}

Finally, defining 
\be
U = e^{\xi-\tau} \qquad,\qquad V = e^{-\xi-\tau}
\ee
gives
\be
ds^2 \sim e^{-2\tau} \left [ d\tau^2 - d\xi^2 - \sinh^2\xi d\Omega_p^2 \right]
- C_0^2 d\Omega_n^2
\ee
as the analytically continued metric just interior to the horizon.

This shows that the extension across the horizon is (cf.\ the global
vortex \cite{GS})
\be
ds^2 = {\bar B}^2(\tau) d\tau^2 - {\bar A}^2 (\tau) dH_{p+1}^2
- {\bar C}^2(\tau) d\Omega_n^2
\ee
where $dH_{p+1}^2$ is the metric on a unit $(p+1)$-dimensional hyperbolic
space. This is of course a time dependent metric as one would expect
for an interior horizon r\'egime.

Following through the computation of the Einstein equations and setting
${\bar B} = {\bar A}$, we once again obtain (\ref{xprime},\ref{yprime})
as the dynamical system for the interior horizon where prime now denotes
$d/d\tau$, and $X$ and $Y$ are defined in terms of the barred variables.
The only difference is that the constraint (\ref{constr}) now reads
\be
{\cal C} (X,Y) = -n(n-1) {{\bar A}^2\over {\bar C}^2} \;.
\ee
\ie\ the interior horizon r\'egime corresponds to the connected
region of the phase plane between the two branches of the invariant
hyperboloid. It is not difficult to verify that the trajectory
from $P_+$ to large negative $X$ and $Y$  in fact corresponds to
an interior horizon solution terminating on a spacelike singularity.

This shows that the causal structure of the inflating $p$-brane is
indeed given by figure \ref{fig:krs}, and the inflating brane
is now a genuine black hole with an horizon. Of course, because
we have only derived general properties of the solution, and
particular asymptotic forms, we do not know the precise value
of the mass of the black hole. Indeed, defining the mass of such
an ALF spacetime is problematic \cite{NS}.

\section{Discussion}

For completeness we would like to remark on the adS $p$-braneworlds.
These are braneworlds in which the metric is a warped product of an
anti de-Sitter braneworld with an $n+1$-dimensional orthogonal space,
and correspond to the dynamical system (\ref{xprime},\ref{yprime}) with
$\kappa = -1$. For this value of $\kappa$ there are no finite critical
points on the phase plane, and the invariant hyperboloid ${\cal C}=0$
now changes from being `timelike' to `spacelike' in the $X-Y$ plane.
The typical trajectory now asymptotes the invariant hyperboloid
at both ends, \ie\ the solution has a null singularity both at
the `core' of the brane, and at finite radial distance. 
Vacuum adS branes are therefore generally singular.
\FIGURE{
\includegraphics[height=15cm]{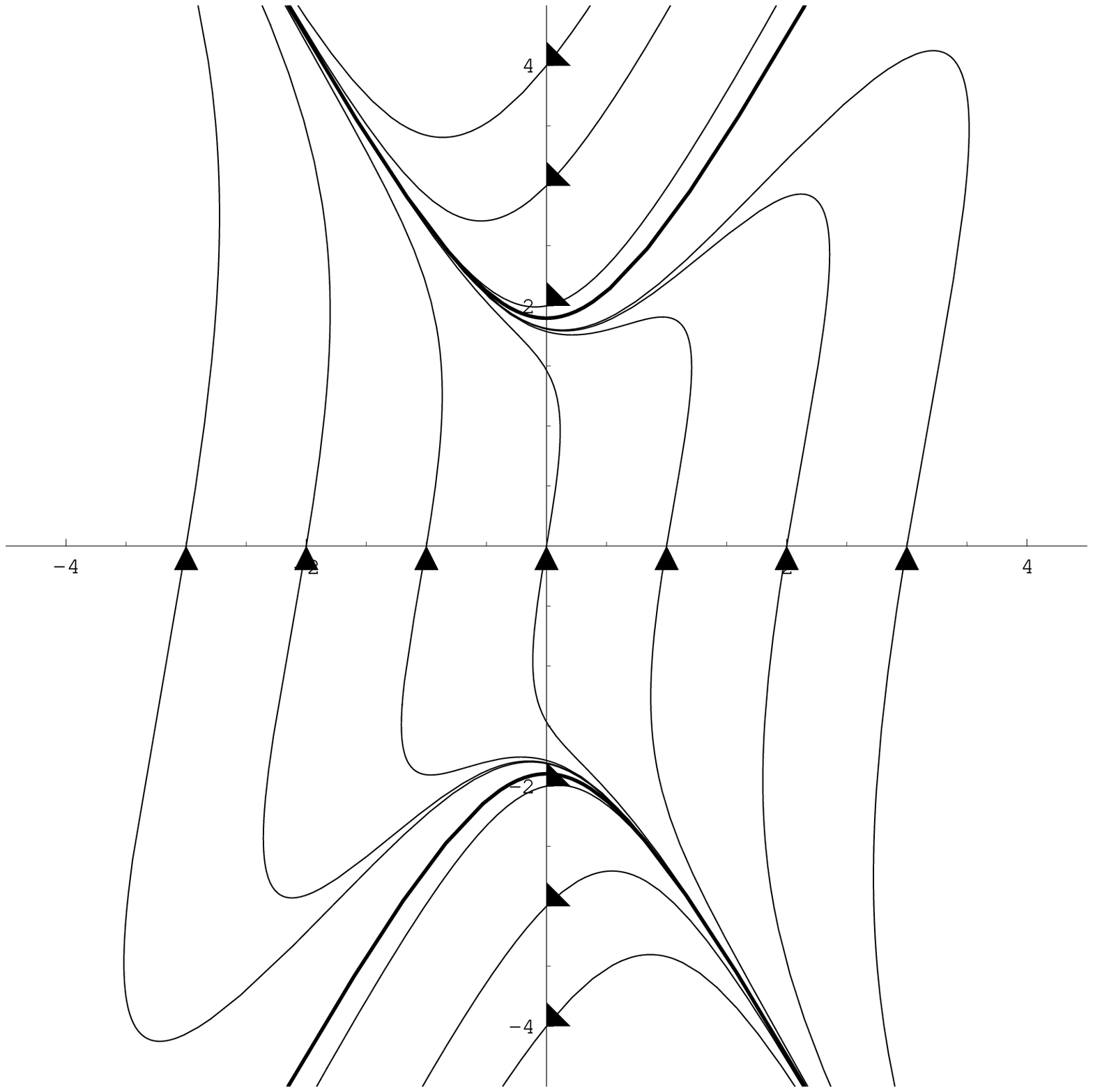}
\caption{The adS brane phase plane. The invariant hyperboloid
${\cal C}=0$ is shown in bold again, but now there are no
critical points.} 
\label{fig:ads32}
}

This result seems in keeping with the general pattern
for the global vortex, \cite{G1}, where one could prove that
the metric was singular for Poincar\'e or adS branes,
but that for a de Sitter geometry on the brane with a fine
tuned Hubble constant related to the brane tension, there was
a nonsingular solution. Here there will be a similar tuning, since 
(\ref{pbrane}) has explicitly set the Hubble constant $H=1$.
We can reintroduce it at the expense of rescaling $r$, and hence
the phase plane.

Of course, this general similarity with the global vortex then
begs the question of whether we can remove the singularity of the
Poincar\'e brane by adding a negative cosmological constant in
the bulk as with the global vortex RS compactification \cite{G2}.
However, a quick look at (\ref{rtt}) shows that this is not
possible. For codimension three or higher, we expect that the null
singularity of the $p$-brane would be smoothed into a horizon
coordinate singularity, \ie\ $A \sim \rho$, $B=1$, $C\sim C_0$
for $\rho\to0$. This is clearly not compatible with (\ref{rtt}).

We can now return to some of the potential applications of 
these inflating solutions and remark on whether they seem
consistent with the assumptions made. First of all, if one
wishes to place branes on compact manifolds, the issue of
the global deficit angles must be addressed. Of course, these
are uncharged branes, however, in \cite{CPB} it was shown that
the general properties of the uncharged spacetime were maintained
when charge and even a dilaton with arbitrary
coupling was also included. The only nonsingular spacetime in that
case was the extremally charged one. Based on general expectation,
and on the results of Berglund et.\ al.\ \cite{DSS}, these would
be expected to actually become singular if one attempted to
introduce inflation on the brane. 

The other interesting application of inflating brane solutions is in
a possible answer to the smallness of the cosmological constant
\cite{Dlam}. In these papers, the authors supposed that introducing
an expansion on the brane would smooth out the null singularity. 
To some extent we have backed up this assumption, however, this
solution is not asymptotically flat, and it is not clear how these
different asymptotics would affect gravity on the brane. Another
key issue is that the authors claim that the Hubble constant is
{\it inversely} proportional to the mass of the brane -- a claim
queried by Cho and Vilenkin \cite{CV}. Questions of how to define
the mass of the brane notwithstanding, we cannot directly comment
on this issue in the absence of an actual solution which would
directly interpolate between the near-horizon solution
(\ref{nrpmet}) and the asymptotic spacetime (\ref{qplussol}).
However, it is interesting to note that increasing the mass of
the Poincar\'e solution actually shrinks the phase plane, in other
words, the trajectories in figure \ref{fig:poin32} with higher mass
are those {\it closer} to the origin. Similarly, introducing an
explicit Hubble expansion on the brane changes the scale of
figure \ref{fig:ds32} moving the critical points $P_\pm$ to
$(\pm H,0)$. Therefore decreasing $H$ also shrinks the phase plane.
Curiously therefore, this scaling does not seem to contradict the
claims of Dvali et.\ al., however, the argument is extremely
unreliable given that $r_+$ tracks a genuine ADM mass in the
Poincar\'e solution, and the thorny issue of mass in the inflating
brane solution needs to be resolved.

To sum up, the pure gravitating $p$-brane can have a nonsingular
(exterior to the horizon) geometry in which the induced metric
parallel to the brane is an inflating de Sitter universe. 
The metric is ALF, but has global deficit angles, and in the 
absence of a core model for the brane has a black hole horizon. 
The presence of the regular horizon, as well as $g_{\theta\theta}$ 
being monotonic for this solution, indicates that we can replace 
this horizon by a general core model in an analogous fashion to 
the replacement of the Reissner-Nordstrom horizon by an SU(2)
monopole core \cite{RNSU2}.

\section*{\bf Acknowledgements.}
 
I would like to thank Christos Charmousis, Gia Dvali, Roberto Emparan, and
particularly Inyong Cho and Alex Vilenkin for discussing their
work with me.
This work was supported by the Royal Society, and I would like to 
acknowledge the Aspen Center for Physics where this work was started.

\end {document}